\documentclass[a4paper,12pt]{article}
\pdfoutput=1    

\usepackage[]{hyperref} 

\usepackage[utf8]{inputenc}
\usepackage{amsmath}
\usepackage{amsthm}
\usepackage{amssymb}
\usepackage{mathrsfs}
\usepackage{amsfonts}
\usepackage{bbold}

\usepackage{algorithm}
\usepackage{algorithmicx}
\usepackage{algpseudocode}
\usepackage{graphicx}
\usepackage{xifthen}
\usepackage{rotating}
\usepackage{xcolor}
\usepackage{subfig} 
\usepackage{psfrag}
\usepackage{amsopn}

\usepackage{authblk}

\usepackage{mathptmx}       
\usepackage{helvet}         
\usepackage{courier}        
\usepackage{type1cm}        
\usepackage{makeidx}         
\usepackage{graphicx}        
\usepackage{multicol}        
\usepackage[bottom]{footmisc}

\usepackage{ucs}              

\usepackage{latexsym}
\usepackage{amsmath,amssymb,amscd,amsbsy}
\usepackage{upgreek}

\usepackage{tensor}

\usepackage{makeidx}         
\usepackage{graphicx}        
\usepackage{multicol}        
\usepackage[bottom]{footmisc}

\usepackage{rcs} \RCS $Revision: 2.2 $ \RCS $Date: 2018/09/02 11:16:02 $ \RCS $Author: hgm $ \RCS $RCSfile: 18_RV-algebra-model.tex,v $ \RCS $Id: 18_RV-algebra-model.tex,v 2.2 2018/09/02 11:16:02 hgm Exp $

\usepackage{anyfontsize}

\usepackage{mathrsfs}
\usepackage{bbm}
\usepackage{amsfonts}
\usepackage{amsmath}
\usepackage{enumerate}

\newcommand{\ignore}[1]{}

\newcommand{\vek}[1]{\mathchoice{\displaystyle\boldsymbol{#1}}
{\textstyle\boldsymbol{#1}}{\scriptstyle\boldsymbol{#1}}
{\scriptscriptstyle\boldsymbol{#1}}}

\newcommand{\tnb}[1]{\mathchoice{\displaystyle\mathboldsans{#1}}
{\textstyle\mathboldsans{#1}}{\scriptstyle\mathboldsans{#1}}
{\scriptscriptstyle\mathboldsans{#1}}}

\usepackage[sfdefault=cmbr,OMLmathsans]{isomath}  
\usepackage{latexsym}
\usepackage{amsmath}
\usepackage{amsthm}
\usepackage{amssymb}
\usepackage{upgreek}  
\usepackage{tensor}  

\usepackage{float}
\usepackage{xcolor}

\makeindex             

\begin{document}


\title{Parameter Identification in Viscoplasticity using Transitional Markov Chain Monte Carlo Method}
\author{{\small Ehsan Adeli and Hermann G. Matthies}}
\affil{{\small Institute of Scientific Computing\\ 
Technische Universit{\"a}t Braunschweig\\
Braunschweig, Germany\\ 
e.adeli@tu-braunschweig.de}}
\date{June 2019}


%
%


\maketitle

\abstract{To evaluate the cyclic behavior under different loading conditions using the kinematic and isotropic hardening theory of steel, a Chaboche viscoplastic material model is employed. The parameters of a constitutive model are usually identified by minimization of the difference between model response and experimental data. However, measurement errors and differences in the specimens lead to deviations in the determined parameters. In this article, the Choboche model is used and a stochastic simulation technique is applied to generate artificial data which exhibit the same stochastic behavior as experimental data. Then the model parameters are identified by applying an estimation using Bayes's theorem. The Transitional Markov Chain Monte Carlo method (TMCMC) is introduced and employed to estimate the model parameters in the Bayesian setting. The uniform distributions of the parameters representing their priors are considered which literally means no knowledge of the parameters is available. Identified parameters are compared with the true parameters in the simulation, and the efficiency of the identification method is discussed. In fact, the main purpose of this study is to observe the possibility of identifying the model and hardening parameters of a viscoplastic model as a very high non-linear model with only a surface displacement measurement vector in the Bayesian setting using TMCMC and evaluate the number of measurements needed for a very acceptable estimation of the uncertain parameters of the model.}

\section{Introduction}
\label{sec:Introduction}
In order to predict the behavior of mechanically loaded metallic materials, constitutive models are applied, which present a mathematical frame for the description of elastic and inelastic deformation. The models by Miller, Krempl, Korhonen,  Aubertin, Chan, and Bodner are well-known constitutive models for isotropic materials \cite{Miller, Krempl, Korhonen1, Aubertin, Chan} which describe the inelastic behavior. In 1983, Chaboche \cite{Chaboche, Chaboche1} put forward what has become known as the unified Chaboche viscoplasticity constitutive model, which has been widely accepted.
	
All inelastic constitutive models contain parameters which have to be identified for a given material from experiments. In the literature only few investigations can be found dealing with identification problems using stochastic approaches. Klosowski and Mleczek have applied the least-squares method in the Marquardt-Levenberg variant to estimate the parameters of an inelastic model \cite{KLOSOWSKI}. Gong et al. have also used some modification of the least-squares method to identify the parameters \cite{Gong}. Harth and Lehn identified the model parameters of a model by employing some generated artificial data instead of experimental data using a stochastic technique \cite{Harth}. A similar study by Harth and Lehn has been done for other constitutive models like Lindholm and Chan \cite{Lindholm}.

There are few investigations on the simplest material model, the elasticity model, to identify only very few parameters of the model where sampling approaches like the Metropolis algorithm and its modifications are employed. Pacheco et al. \cite{Pacheco} investigated a three-point bending test to identify elastic behavior and calibration is done by solving the inverse problem in a Bayesian setting by using the Metropolis-Hastings algorithm. Only elastic moduli are identified without considering the error. Slonski et al. \cite{Marek} also applied a sequential particle filter on an elastic model and Young’s modulus is estimated, where the Bayesian setting is compared to the deterministic approach, and the Bayesian setting is preferred. The elastic modulus of a polymeric material is updated by Zhang et al. \cite{Zhang}, where a Markov Chain Monte Carlo method is used, but very high computation time is reported. 
Further, Young’s modulus is estimated for a considered material by Gallina et al. \cite{Gallina1} by applying a multi dimensional Markov Chain Monte Carlo method. A similar Bayesian approach for composite materials to estimate Young’s modulus is carried out by Pieczonka et al. \cite{Pieczonka}. 
Arnst et al. \cite{Arnst} have applied a Markov Chain Monte Carlo method by using polynomial chaos expansion to identify Young’s modulus of an elastic model.

Only few investigations were carried out on enriched material models such as a viscoelastic model to identify the few parameters by employing the Metropolis-Hastings technique and the classical Markov Chain Monte Carlo method. Rappel et al. \cite{Rappel, Rappel1} studied an elastic and a viscoelastic model where the measurement error is considered. Bayesian inference is applied to estimate only the elasticity modulus by applying an adaptive Metropolis-Hastings technique. An et al. \cite{An} investigated a crack model by a classical Markov Chain Monte Carlo method in order to estimate the parameters of a model which represent the size and position of the crack in the Bayesian setting. Also Hernandez et al. \cite{Hernandez} applied a Markov Chain Monte Carlo method on a viscoelastic model in order to update its model parameters in a Bayesian setting, but the posterior distributions of the parameters are not updated properly. In fact, the parameters are not identified properly. 
Mahnken \cite{Mahnken} has also applied a Markov Chain Monte Carlo method to estimate few parameters of a plasticity model.
The damage parameters of a truss structure under model uncertainties are studied by Zheng et al. \cite{Zheng}, where a multi-level Markov Chain Monte Carlo method is applied and the true values are not well estimated. Further, this approach suffers from high computation time. Another damage detection approach is applied by Nichols et al. \cite{Nichols} by applying a modified version of the Markov Chain Monte Carlo method.

There are other investigations in the Bayesian setting using the Markov Chain Monte Carlo method, and Madireddy et al \cite{Madireddy1}, Wang and Zabaras \cite{Zabaras}, and Oh et al \cite{Oh} have carried out an investigation on the identification of a material model by using its modified method. In studies carried out by Alvin \cite{Alvin}, Marwala and Sibusiso \cite{Marwala1}, Daghia et al. \cite{Daghia}, Abhinav and Manohar \cite{Abhinav}, Gogu et al. \cite{Gogu, Gogu1}, and Koutsourelakis \cite{Koutsourelakis, Koutsourelakis1}, the elastic parameters of the model are estimated stochastically. Fitzenz et al. \cite{Fitzenz}, Most \cite{Most1}, and Sarkar et al. \cite{Sarkar} investigated elastoplastic materials and thermodynamical material models to identify their model parameters. Other studies on viscoelastic models are carried out by Zhang et al. \cite{Zhang1}, Mehrez et al. \cite{Mehrez}, and Miles et al. \cite{Miles}. Further investigations on viscoelastic models to estimate a higher number of model parameters are studied by Zhao and Pelegri \cite{Zhao123}, and by Kenz et al. \cite{Kenz}. 
The estimation of fatigue parameters using Markov Chain Monte Carlo methods is also studied by \cite{Dan}.

In this paper, a viscoplastic model of Chaboche is studied. The model contains five material parameters which have to be determined from experimental data. It should be noted that here virtual data are employed instead of real experimental data. A cyclic tension-compression test is applied in order to extract the virtual data.

The model is described in Section 2, whereas Section 3 explains how to perform the update. The employed method, which is the Transitional Markov Chain Monte Carlo (TMCMC) method, is described in this section and applied on the case study model.

In Section 4 the desired parameters of the model are identified from the measured data which .. take as surface displacement measurements, as could be done for example by digital image correlation (DIC). In fact, the parameters which have been considered as uncertain parameters are updated and the uncertainties of them are reduced, while the random variables representing the uncertain parameters are updated during the process. The results are thoroughly studied and the identified parameters as well as the corresponding model responses are analyzed. The section discusses the ability of the applied method to identify the highly non-linear viscoplastic model parameters by considering the surface displacement measurement.

\section{Model Problem}
The mathematical description of metals under cyclic loading beyond the yield limit that includes viscoplastic material behavior as well as the characterization of isotropic-kinematic hardening is here given in terms of a modified Chaboche model introduced in \cite{dinkler}. As we consider classical infinitesimal strains, we assume an additive strain decomposition of elastic and inelastic strains. The material behavior is described for the elastic part by isotropic homogeneous elasticity which is characterized by the shear and bulk moduli, the eigenvalues of the elasticity tensor. The inelastic behavior is described following the standard generalized materials paradigm \cite{quoc}, so that obey the specification of the dissipation pseudo-potential is necessary. For viscoplasticity the dissipation potential is given by
\begin{equation}
     \phi(\vek{\sigma}) = \frac{k}{n+1}\Biggl\langle\frac{\sigma_{ex}}{k}\Biggr\rangle^{n+1},
     \label{eq:5}
\end{equation}
with $\langle\cdot\rangle = \max(0, \cdot)$ and $k$ and $n$ as the material parameters. Here $\sigma_{ex}$ is the over-stress, defined via the equivalent stress ($\sigma_{eq}$) which reads
\begin{equation}
\sigma_{eq}= \sqrt{\frac{3}{2} \text{tr} ((\vek{\sigma}-\vek{\chi})_D . ((\vek{\sigma}-\vek{\chi})_D)},
\end{equation}
where $(\cdot)_D$ denotes the deviatoric part and $\vek{\chi}$ is the back-stress of kinematic hardening. The over-stress $\sigma_{ex}$ is given by
\begin{equation}
    \sigma_{ex} = \sigma_{eq}-\sigma_y -R= \sqrt{\frac{3}{2} \text{tr} ((\vek{\sigma}-\vek{\chi})_D . (\vek{\sigma}-\vek{\chi})} - \sigma_y - R,
\end{equation}
where $\sigma_y$ is the yield stress and $R$ models the isotropic hardening which is introduced in the following. The partial derivative of the dissipation potential $\phi$ with respect to $\vek{\sigma}$ leads to the equation for the inelastic strain rate
\begin{equation}
    \dot{\vek{\epsilon}}_{vp} = \frac{\partial \phi}{\partial \vek{\sigma}} = \Biggl\langle\frac{\sigma_{ex}}{k}\Biggr\rangle^{n} \frac{\partial \sigma_{ex}}{\partial \vek{\sigma}}.
\end{equation}
The viscoplastic model allows for isotropic and kinematic hardening, which is considered in order to describe different specifications. Assuming $R(t)$ and $\vek{\chi}(t)$ with $R(0) = 0$ and $\vek{\chi}(0) = 0$ to describe isotropic and kinematic hardening respectively, the evolution equations for these two are
\begin{equation}
\dot{R} = b_R(H_R-R)\dot{p}
\end{equation}
and
\begin{equation}
\dot{\vek{\chi}} = b_{\vek{\chi}}(\frac{2}{3} H_{\vek{\chi}}\frac{\partial \sigma_{eq}}{\partial \vek{\sigma}} - \vek{\chi} )\dot{p}
\end{equation}
respectively. In the evolution equations, $\dot{p}$ is the viscoplastic multiplier rate given as:
\begin{equation}
    \dot{p} = \Biggl\langle\frac{\sigma_{ex}}{k}\Biggr\rangle^{n},
\end{equation}
which describes the rate of accumulated plastic strains. The parameter $b_R$ indicates the speed of stabilization, whereas the value of the parameter $H_R$ is an asymptotic value according to the evolution of the isotropic hardening. Similarly, the parameter $b_{\vek{\chi}}$ denotes the speed of saturation and the parameter $H_{\vek{\chi}}$ is the asymptotic value of the kinematic hardening variables. The complete model is stated in Table~\ref{tab:1}. Note that $\tnb{E}$ represents the elasticity tensor.

\begin{table}[H]
\caption{The constitutive model of Chaboche}
\begin{center}
\begin{minipage}{0.8\textwidth}
 \begin{tabular}{|| l||}   
 \hline
 \label{tab:1}
 Strain\\
 \centerline{$\vek{\epsilon}(t) = \vek{\epsilon}_e(t) + \vek{\epsilon}_{vp}(t)$} \\ \\
 Hooke's Law\\
 \centerline{$\vek{\sigma}(t) = \tnb{E}: \vek{\epsilon}_e(t)$}\\ \\
 Flow Rule\\
 \centerline{$\dot{\vek{\epsilon}}_{vp}(t) = 
     \langle\frac{\sigma_{eq}(t)-\sigma_y-R(t)}{k}\rangle^n \frac{\partial \sigma_{ex}}{\partial \vek{\sigma}}$}  \\ \\
 Hardening\\
  \centerline{$\dot{R} = b_R(H_R-R)\dot{p}$}\\ \\
 \centerline{$\dot{\vek{\chi}} = b_{\vek{\chi}}(\frac{2}{3} H_{\vek{\chi}}\frac{\partial \sigma_{eq}}{\partial \vek{\sigma}} - \vek{\chi} )\dot{p}$}\\ \\

 Initial Conditions\\
  \centerline{$\vek{\epsilon}_{vp}(0)=0$,~~$R(0)=0$,~~$\vek{\chi}(0)=0$}\\ \\
  
 Parameters\\
   \centerline{ $\sigma_y$ ~(Yield Stress)}\\ 
   \centerline{ $k$, $n$ ~(Flow Rule)}\\  
   \centerline{ $b_R$, $H_R$, $b_{\vek{\chi}}$, $H_{\vek{\chi}}$~(Hardening)}\\
 [1ex] 
 \hline
\end{tabular}
\end{minipage}
\end{center} 
\end{table}

By gathering all the desired material parameters to identify into the vector $\vek{q}=[\kappa~ G~ b_R~ b_{\vek{\chi}}~ \sigma_y]$, where $\kappa$ and $G$ are bulk and shear modulus, respectively, which determine the isotropic elasticity tensor, the goal is to estimate $\vek{q}$ given measurement displacement data, i.e. 
\begin{equation}
u=Y(\vek{q})+e
\label{eq:pce_uf_yf},
\end{equation}
in which $Y(\vek{q})$ represents the measurement operator giving the prediction of the observed displacements and $e$ is the measurement (also possibly the model) error. As the function $Y(\vek{q})$ is typically not invertible, this is an ill-posed problem, and the estimation of $\vek{q}$ given $u$ is not an easy task and usually requires regularization. This can be achieved either in a deterministic or in a probabilistic setting. Here, the latter path is taken as further described in the text.

\section{Bayesian identification via the Transitional Markov Chain Monte Carlo method}
Ching and Chen \cite{Ching} modified the approach by Beck and Au to develop a Transitional Markov Chain Monte Carlo method (TMCMC). This approach is based on the Markov Chain Monte Carlo method and its adaptive capability is inspired from an adaptive Metropolis–Hastings (M-H) method developed by Beck and Au in 2002 \cite{SK4}. This approach employs a sequence of PDFs, where the entire data set of the available data is used for each stage of the sampler which is proportional to the local posterior with some exponent times likelihood, with a change of exponent between 0 and 1. As the same idea was used in the simulated annealing approach \cite{Fishman}, this exponent is called a tempering parameter. TMCMC and the Beck and Au approach also differs in theory of employing the M-H algorithm. In each stage a local proposed PDF is constructed instead of a global proposed PDF, and it is also to be noted that to improve the rate of convergence in TMCMC, re-sampling is used.

Based on the formulation of Bayes's theorem \cite{Aster, Dashti}, it is difficult to generate samples from the posterior PDF due to lack of information about the geometry of the probability density function. In order to overcome this problem, the TMCMC algorithm employs a sequence of intermediate PDFs which converge to the target posterior PDF as defined in the Equation \ref{eq:Ching posterior PDF}, where the index $j$ denotes the stage number \cite{Ching, Ching1}. Considering the reformulation of Bayes's theorem for the model class $\mathcal{M}$ which is defined by the parameter vector $\vek{\theta}$, the prior PDF of this model class over the parameter vector $f(\vek{\theta}| \mathcal{M})$ is updated with the measurement data $\mathcal{D}$, where the likelihood function is $f(\mathcal{D} |  \mathcal{M}, \vek{\theta})$.

\begin{subequations}
\label{eq:Ching posterior PDF}
\begin{align}
&
  f_j(\vek{\theta}) \propto f(\vek{\theta}|\mathcal{M})f(\mathcal{D}|\mathcal{M},\vek{\theta})^{r_j}    \\ 
    j = 0,1,2,...,& M  \,\,\,\,\,\,\,\,\,\,\,\,\,\,  0=r_0 \le r_1 \le r_2 \le ... \le r_M =1   
\end{align}
\end{subequations}

The stage number fulfills the desirable properties such as producing a series of intermediate PDFs ($f_j(\vek{\theta})$) from the prior PDF, i.e. it starts with $r_j = 0$ stating that the initial prior is proportional to the prior PDF ($f_0(\vek{\theta}) \propto f(\vek{\theta}|\mathcal{M})$) and it ends with $r_j = 1$ stating that the final PDF is proportional to the posterior PDF ($f_M(\vek{\theta}) \propto f(\vek{\theta}|\mathcal{M})f(\mathcal{D}|\mathcal{M},\vek{\theta})$). The right hand side of the latter relation represents the posterior PDF $f(\vek{\theta}|\mathcal{M},\mathcal{D})$ and its proportionality is shown in Equation \ref{eq:Ching desirable properties}.

\begin{subequations}
\label{eq:Ching desirable properties}
\begin{align}
&
  r_j = 0 \,\,\,\,\,\,\,\,\,\,\,\,\,\, f_0(\vek{\theta}) \propto f(\theta|\mathcal{M})    \\ 
    r_j & = 1 \,\,\,\,\,\,\,\,\,\,\,\,\,\, f_M(\vek{\theta}) \propto f(\vek{\theta}|\mathcal{M},\mathcal{D})   
\end{align}
\end{subequations}

Although the geometry change from $f(\vek{\theta}|\mathcal{M})$ to $f(\vek{\theta}|\mathcal{M},\mathcal{D})$ is large, the change between two adjacent intermediate PDFs can be small. Due to this possible transition, from where the method derives its name, it is possible to determine samples efficiently from $f_{j+1}(\vek{\theta})$ based on samples from $f_{j}(\vek{\theta})$.  


\subsection{Transitional MCMC Algorithm}

In this subsection, the entire algorithm of TMCMC is described in the following steps in detail \cite{Ching, Ching1, Jia}.

\begin{enumerate}
 \item 
 As a first step, set $j=0$, and generate from $f_0(\vek{\theta}) = f(\vek{\theta}|\mathcal{M})$ a set of $N$ samples.
 \item 
 It is to be noted that $r_{j+1}$ is selected in such a way that the coefficient of variation of $f(\mathcal{D}|\mathcal{M},{\vek\theta}_k^{(j)})^{r_{j+1}-r_j}$ where $k=1,2,...,N$ is a prescribed value and $\vek{\theta}_k^{(j)}$ stands for the $k$-th sample that belongs to level $j$. Accordingly, the coefficient of variation serves as an indicator to measure the closeness of $f_j(\vek{\theta})$ to $f_{j+1}(\vek{\theta})$.
 \item 
 The plausibility weight $w(\vek{\theta}_k^{(j)})=f(\mathcal{D}|\mathcal{M},\vek{\theta}_k^{(j)})^{r_{j+1}-r_j}$ is obtained for $k=1,2,...,N$, from which the parameter $S^j=\sum_{k=1}^{N}w(\vek{\theta}_k^{(j)})/N^j$ is computed.
 \item 
 Samples $\vek{\theta}_k^{j+1}$ where $k=1,2,...,N$ determined from $f(\vek{\theta})^{j+1}$ by Metropolis-Hastings technique are chosen randomly from a Markov chain which has samples starting from one of the samples $\vek{\theta}_i^{j}$ where $i=1,2,...,N$. The i-th initial sample $\vek{\theta}_i^{(j)}$ is chosen with probability $w(\vek{\theta}_i^{(j)})/\sum_{l=1}^{N}w(\vek{\theta}_l^{(j)})$. As in the proposed distribution a Gaussian distribution which is centered at the current sample is applied in the k-th chain in the Metropolis-Hastings algorithm, and the determined covariance matrix is shown in Equation \ref{eq:covariance matrix} and \ref{eq:scaling factor}, where $\beta$ is a prescribed scaling factor 
 that scales the proposal distribution and $\beta = 0.2$ is suggested by Ching and Chen \cite{Ching}.
 \newline
 \begin{equation}
  \tnb{\Sigma}_{j} = \beta^2 \frac{\sum_{i=1}^{N}w(\vek{\theta}_i^{(j)})\left[\vek{\theta}_i^{(j)}-\vek{\mu}_j\right]\left[\vek{\theta}_i^{(j)}-\vek{\mu}_j\right]^T } {\sum_{k=1}^{N}w(\vek{\theta}_k^{(j)})}
  \label{eq:covariance matrix}
 \end{equation}
 where 
 \newline
  \begin{equation}
   \vek{\mu}_{j} = \frac{\sum_{l=1}^{N}w(\vek{\theta}_l^{(j)}) \vek{\theta}_l^{(j)}} {\sum_{l=1}^{N}w(\vek{\theta}_l^{(j)})}.
   \label{eq:scaling factor}
  \end{equation}
  \newline
 \item 
 Steps two to four are repeated until $r_M=1$. At the final step, samples $\vek{\theta}_k^{(M)}$ for $k=1,2,...,N$ are distributed according to $f(\vek{\theta}|\mathcal{M},\mathcal{D})$. 
\end{enumerate}

The significant advantage of the modified Metropolis-Hastings algorithm applied in TMCMC is that initially it allows a large and a free sample space to explore but in the final stages the samples are generated from a narrower neighborhood of sample space. Furthermore the proposal distribution will change within a simulation level and this will result in better local behavior. This is accomplished by modifying the proposal distribution for each level in such a way that its standard deviation is small for higher simulation levels. But the average value drives the generation of samples towards the most important neighborhood of the sample space \cite{Ching, Ching1, Jia}.

\section{Numerical results}

The identification of the material constants in the Chaboche unified viscoplasticity model is a reverse process, here based on virtual data. In case of the Chaboche model the best way of parameter identification is using the results of cyclic tests, since more information can be obtained from cyclic test rather than creep and relaxation tests, specifically information regarding hardening parameters. The aim of the parameter identification is to find a parameter vector $\vek{q}$ introduced in Section 2. The bulk modulus ($\kappa$), the shear modulus ($G$), the isotropic  hardening coefficient ($b_{R}$), the kinematic hardening coefficient ($b_{\vek{\chi}}$) and the yield stress ($\sigma_y$) are considered as the uncertain parameters of the constitutive model. 

\begin{figure}[H]
    \begin{center}{}
      \includegraphics[width=2.5in]{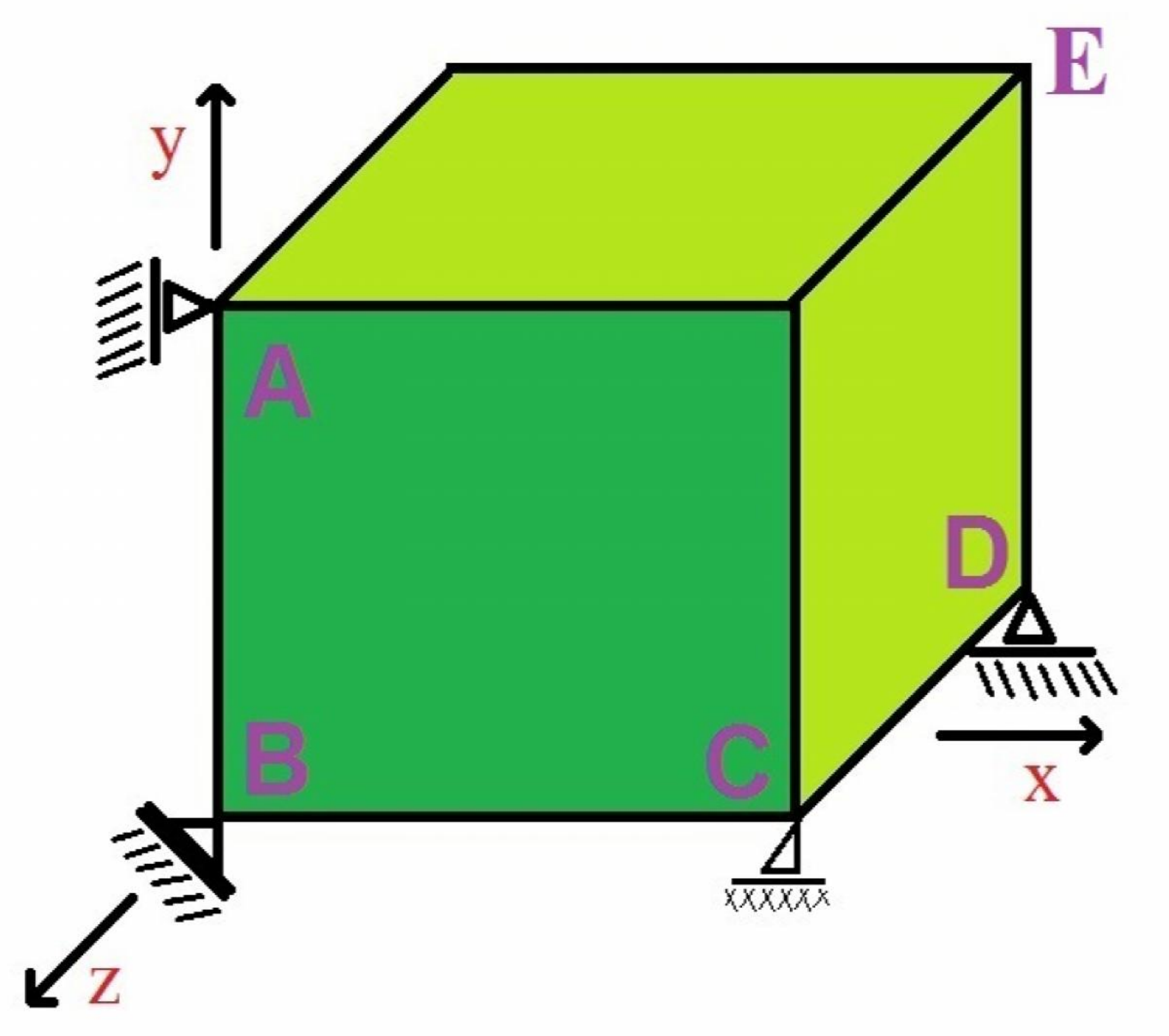}
     \caption{\label{fig:1eleBC6} Boundary condition considered}
    \end{center}\hspace{2pc}%
\end{figure}

A preliminary study is on a regular cube, modeled with one 8 node element. The minimal number of freedoms that have to be constrained is six and many combinations are possible. As shown in Figure \ref{fig:1eleBC6}, all three degrees of freedom at point B are fixed. This prevents all rigid body translations, and leaves three rotations to be taken care of. The $x$ displacement component at point A is constrained to prevent rotation about $z$, and the $z$ component is fixed at point C to prevent rotation about $y$. The $y$ component is constrained at point D to prevent rotation about $x$ \cite{Felippa}.

\begin{figure}[H]
    \begin{center}{}
      \includegraphics[width=3.0in]{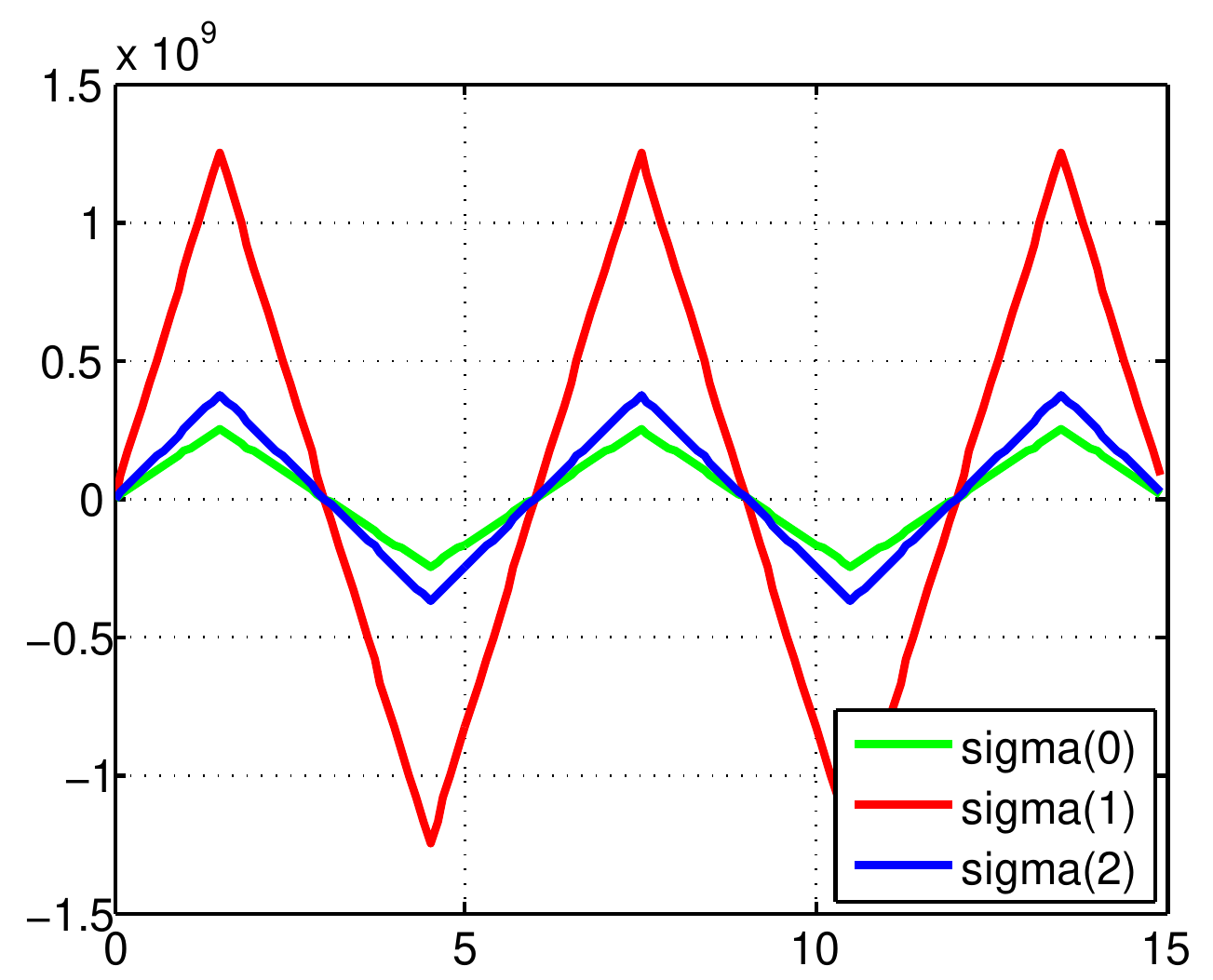}
     \caption{\label{fig:sigma-timeelastic} Decomposed applied force at point E according to time}
    \end{center}\hspace{2pc}%
\end{figure}

The normal tractions, which is a Neumann boundary condition, are applied cyclically \cite{Adeli, Adeli1, Adeli2} in $x$, $y$ and $z$ directions on front and back faces and the magnitude of tractions in all directions are shown in Figure \ref{fig:sigma-timeelastic} where green, red and  blue colors represent the stress vector values in $x$, $y$ and $z$ directions respectively.

\begin{figure}[H]
\centering
\includegraphics[width = 3.0in]{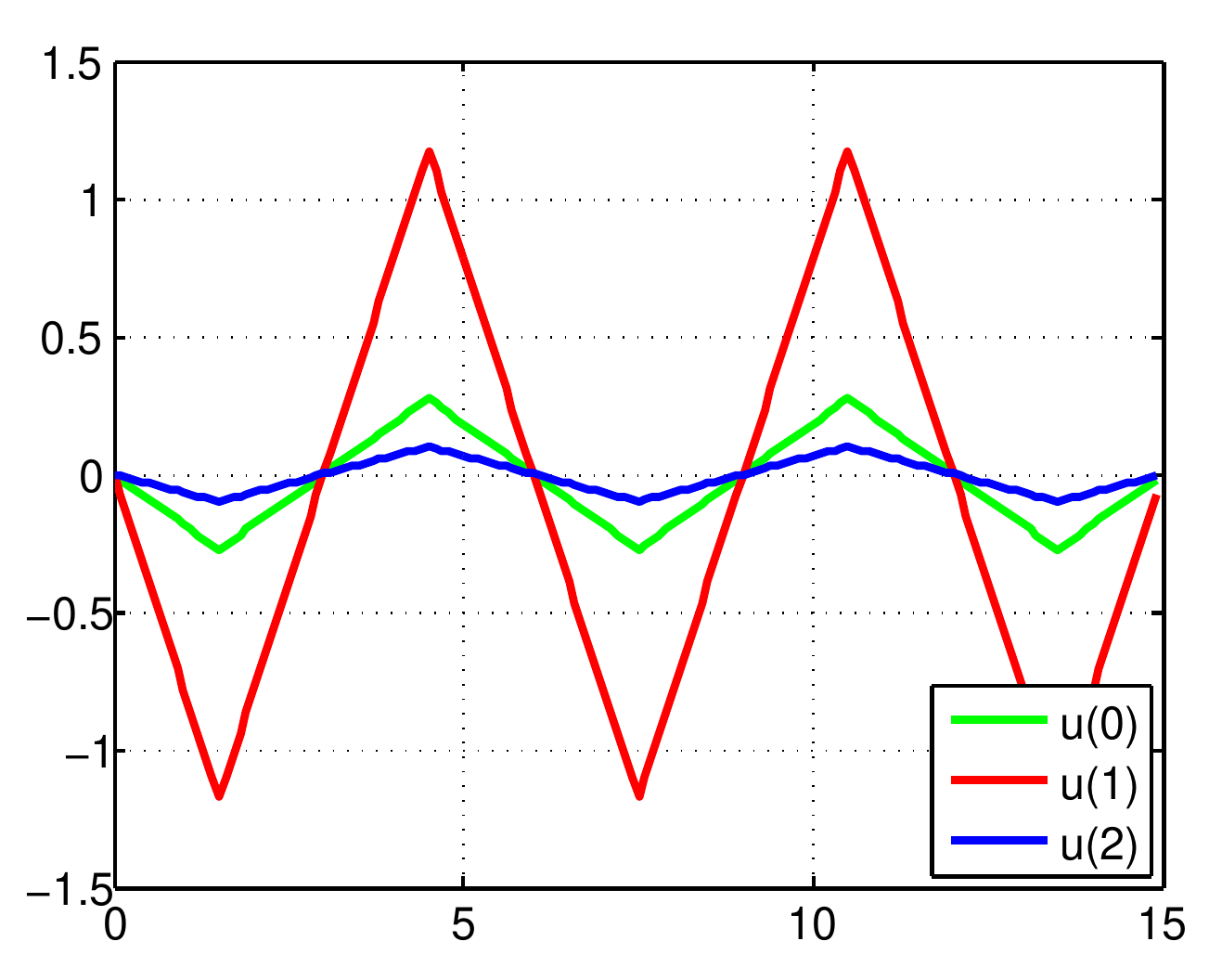}
\caption{\label{fig:sigma-epsilon} Displacement of point E in $x$, $y$ and $z$ directions according to time}
\end{figure}

By considering the parameters listed in Table \ref{tab:elastic1} as the virtual truth, for the top right corner node on back face, point E, as shown in Figure \ref{fig:1eleBC6}, the related displacement graph is obtained as shown in Figure \ref{fig:sigma-epsilon} where green, red and blue colors represent the displacement of point E in $x$, $y$ and $z$ directions respectively.

\begin{table}[H]
\caption{The model parameters}
\begin{center}
\begin{tabular}{l*{9}{c}r}
\hline
              & $\kappa$ & $G$ & $\sigma_y$ & $n$ & $k$ & $b_{R}$ & $H_{R}$ & $b_{\chi}$ &  $H_{\chi}$ &  \\
\hline
             & $1.66\mathrm{e}9$ & $7.69\mathrm{e}8$  &$1.7\mathrm{e}8$ & $1$ & $1.5\mathrm{e}8$ & $50$ & $2.75\mathrm{e}8$ & $50$ &  $2.75\mathrm{e}8$ &  \\
\hline
\label{tab:elastic1}

\end{tabular}
\end{center}
\end{table}

The displacements of point E in $x$, $y$ and $z$ directions are noted as the virtual data in this case. 

\begin{figure}[H]
\centering
\includegraphics[width = 3.75in]{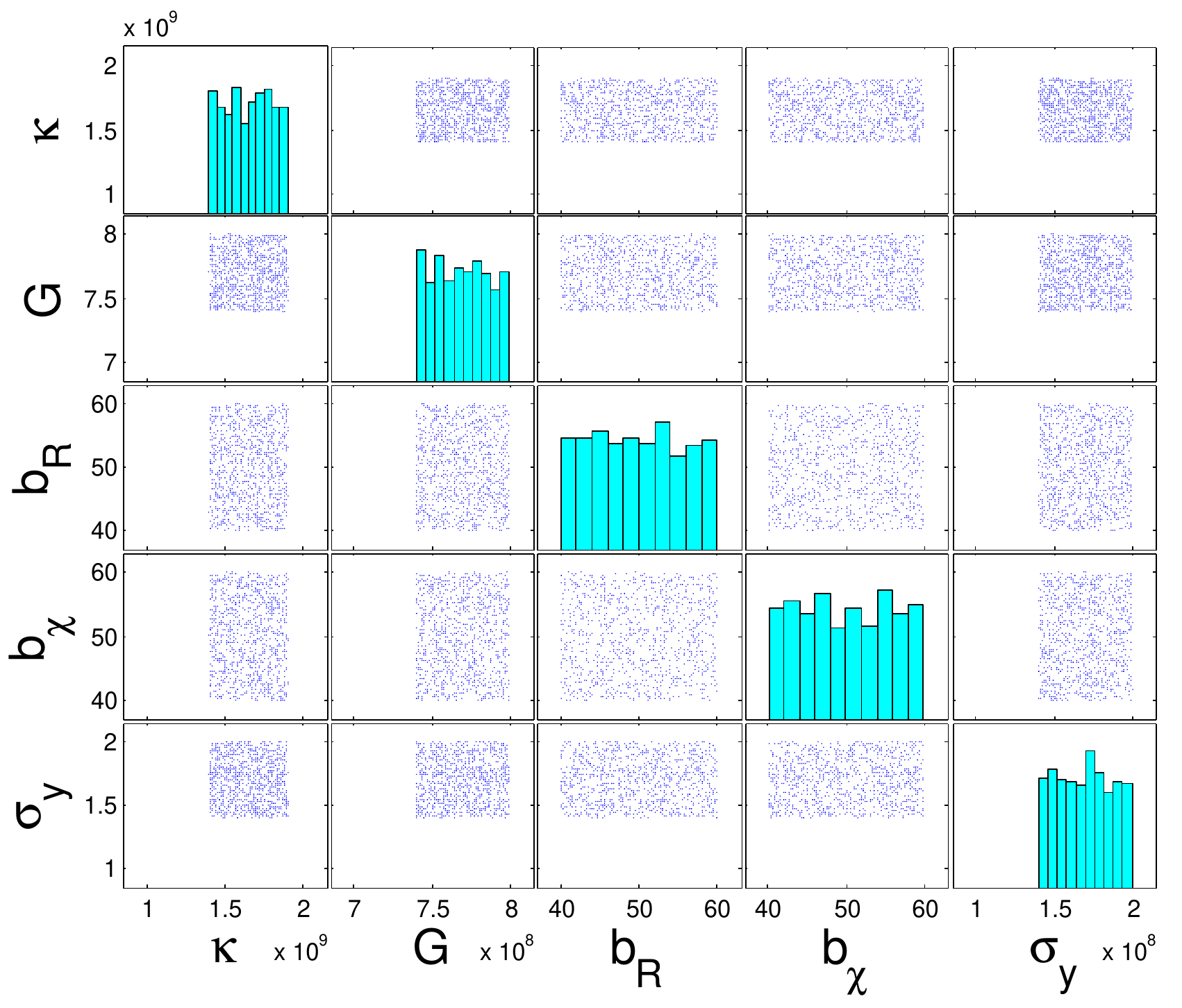}
\caption{\label{fig:elasticTMCMC1} Prior distribution of the parameters}
\includegraphics[width = 3.75in]{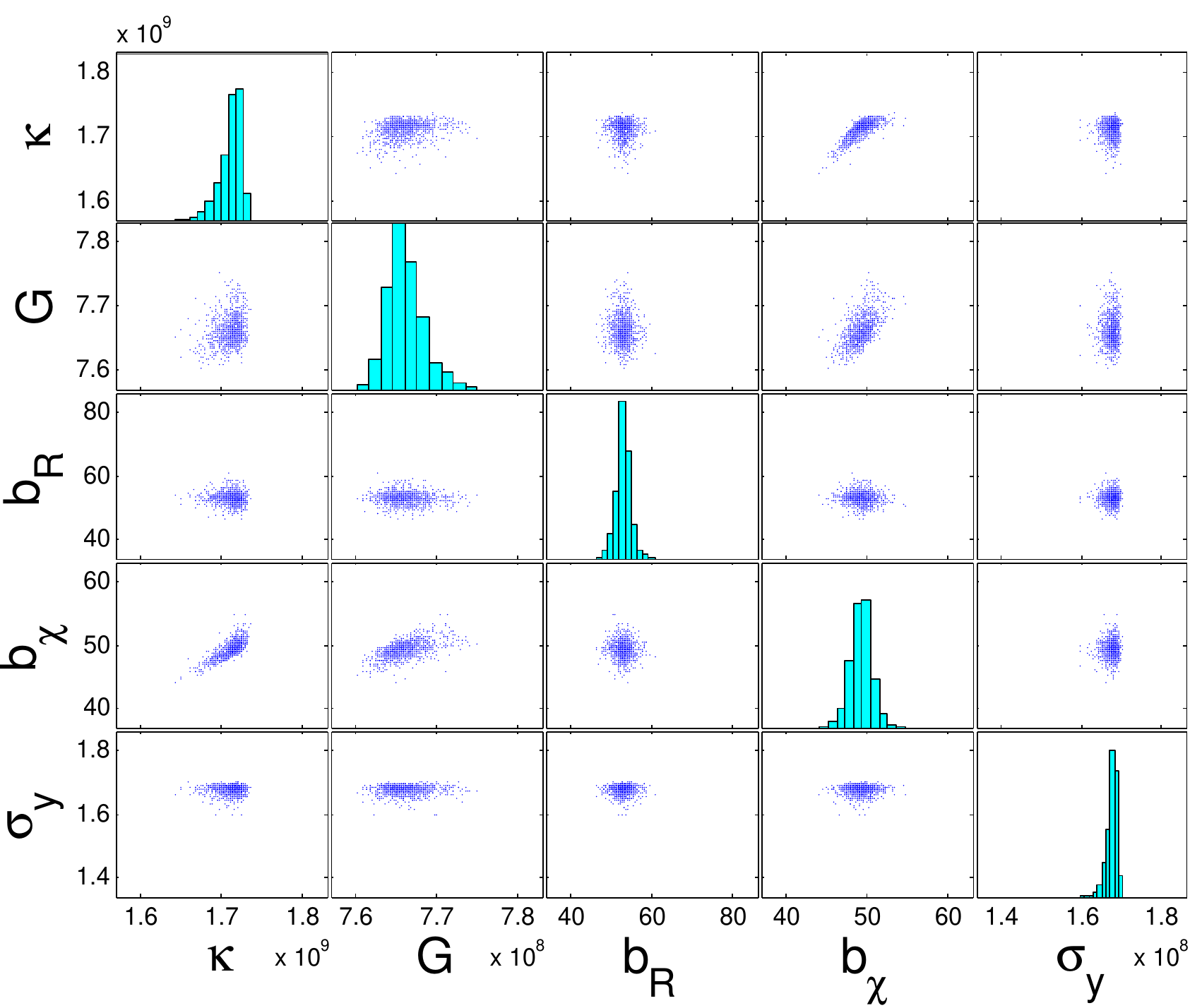}
\caption{\label{fig:elasticTMCMC2} Posterior distribution of the parameters}
\end{figure}

The Transitional Markov chain Monte Carlo method is applied using sampling by which 1000 samples are generated and the history of the displacement of point E is noted. The determined prior and posterior probability density functions are compared and illustrated in Figure \ref{fig:elasticTMCMC1} and \ref{fig:elasticTMCMC2}. It should be noted that the Burn-in period is considered as $200$ in all iterations but in last iteration it is considered as $500$. Further, the scaling parameter ($\beta$) in TMCMC is considered as $0.2$ \cite{Ching}. The acceptance ratio is implemented in such an adaptive way that it is between 0.2 and 0.3, considering the optimal acceptance ratio 0.234 \cite{Gelman}. The initial prior of the material model parameters are taken as uniform distributions. 

From the prior and posterior distributions of the bulk modulus (${\kappa}$), shear modulus ($G$), the isotropic hardening coefficient ($b_{R}$), the kinematic hardening coefficient ($b_{\chi}$) and the yield stress ($\sigma_y$), it is found that the model parameters can be detected by using the TMCMC. The obtained results are acceptable, however it should be mentioned that the TMCMC method is a very time consuming approach, particularly when the case study is a PDE system. The determined true values, the mean and standard deviation of the estimated parameters via the TMCMC approach are shown in Table~\ref{tab:elasticTMCMC}.

\begin{table}[H]
\caption{The identified model parameters}
\begin{center}
\begin{tabular}{l*{3}{c}r}
\hline
Parameters               & $\vek{q}_{\text{true}}$  & $\vek{q}_{\text{est}}(\text{mean})$ & $\vek{q}_{\text{est}}(\text{standard deviation})$ & \\
\hline
${\kappa}$              & $1.66\mathrm{e}9$ &$1.71\mathrm{e}9$ & $1.33\mathrm{e}7$ &  \\
$G$             & $7.69\mathrm{e}8$  &$7.66\mathrm{e}8$ & $2.24\mathrm{e}6$ &  \\
$b_{R}$              & 50  & 53.00 & 1.83 &  \\
$b_{\chi}$             & 50  & 49.30 & 1.32 &  \\
$\sigma_y$                  & $1.7\mathrm{e}8$  & $1.67\mathrm{e}8$ & $1.37\mathrm{e}6$ & \\
\hline
\end{tabular}\label{tab:elasticTMCMC}
\end{center}
\end{table}

\subsection{Discussion of the results}

From the sharpness of the posterior PDF of $\kappa$, $G$, $b_{R}$, $b_{\chi}$ and $\sigma_{y}$, it can be concluded that enough information from virtual data, the surface displacement measurement, is received, and updating the parameters considering their uncertainty is done very properly, as the standard deviation of the residual uncertainty is below $1\%$ of the mean. On the other hand, it should be mentioned that the Transitional Markov Chain Monte Carlo approach, which is employed, is a very time consuming method and applying it, besides a model which is solved by employing the finite element method as a relatively expensive method, leads to a computationally very expensive approach to estimate the model parameters.

\section {Summary}
Using the Transitional Markov Chain Monte Carlo method explained in Section 3 to identify the model parameters of the Chaboche model indicates that it is possible to identify the model parameters by employing this method. The method works well on this highly non-linear material model by taking the vector of surface displacement measurement into consideration. The parameters are well estimated, and the uncertainty of the parameters is reduced, while the distribution functions of the parameters are updated during the process.  The adaptive procedure of the acceptance ratio, where the acceptance ratio is accepted when it is between 0.2 and 0.3, leads to very accurate posterior distributions, which estimate the parameters very finely. As applying the TMCMC on the discussed material model needs a lot of computational time, a new much faster approach will be introduced to estimate the model parameters using the surface displacement measurement and the results will be compared on the same material model in the upcoming paper of the authors which can be proposed not only for the purpose of model parameter identification but also for the purpose of structural health monitoring and damage detection of the real structures such as bridges under cyclic tests.
\\

%
\textbf{\textit{Acknowledgement}} \\
This work is partially supported by the DFG through GRK 2075.
%



\end{document}